\documentclass[aps,preprint,amsmath,amssymb]{revtex4}
\usepackage{graphicx,color}
\def\be{\begin{eqnarray}}
\def\ed{\end{eqnarray}}
\def\non{\nonumber}
\def\ga{\gamma}

\def\l{\lambda}

\def\la{\langle}
\def\ra{\rangle}

\begin{document}

\title{\Large \bf $\Delta m_D$ and $\Delta \Gamma_D$ revisted}

\date{\today}

\author{\bf
Chuan-Hung~Chen$^{1,2}$\footnote{E-mail: physchen@mail.ncku.edu.tw
}, C.~S.~Kim$^{3}$\footnote{E-mail: cskim@yonsei.ac.kr,
}  }
\affiliation{
$^{1}$ Department of Physics, National Cheng-Kung University, Tainan 701, Taiwan \\
$^{2}$National Center for Theoretical Sciences, Hsinchu 300, Taiwan
\\
$^{3}$ Department of Physics $\&$ IPAP, Yonsei University, Seoul 120-479, Korea
}

\begin{abstract}
\noindent The lifetime difference ($y_D=\Delta\Gamma_D/2\Gamma_D$) and mass difference ($x_D=\Delta m_D/\Gamma_D$)  of neutral $D$ meson
have been measured with $y_D=(0.80\pm 0.13)\%$ and $x_D=(0.59 \pm 0.20)\%$, respectively.
Intriguingly,  in contrast with the cases of $K$ and $B_q$ systems,
the current data indicate that $y_D/x_D \sim 1$ and $y_D$ favors to be larger than $x_D$. For explaining the experimental indication,
we here study the $D-\bar D$ oscillation in the framework of unparticle physics.
We demonstrate that {\it the peculiar phase appearing in off-shell unparticle propagator} could play an important role
on $x_D$ and $y_D$.
\end{abstract}

\maketitle

In the Standard Model (SM), the most impressive features of flavor physics are the Glashow-
Iliopoulos-Maiani (GIM) mechanism \cite{Glashow:1970gm} and the large top quark mass.
The former results in the cancelation between the first two generations so that the mass difference $\Delta m_K $
in the neutral $K$ system could be suppressed, while the latter makes $\Delta m_{B_q}$ (q = d, s)
in the $B_q$ systems dominated by the short-distance (SD) top-quark effects \cite{Hagelin:1981zk}.
Due to the precision measurements and the sensitivity to the new physics,
within the past decades  enormous studies have been done in $K$ and $B_q$ mesons,  which are composed of down type quarks.

By the production of large number of $D$ mesons at Tevatron and $B$ factories worldwide,
now the neutral charmed meson which is made up of up-type quarks also
plays an important role on the test of the SM.
By the world average, the current measurements with allowing CP violation (CPV) for $D-\bar D$ mixing  are  given by \cite{HFAG}
 \be
 x_D &=& \frac{ m_H - m_L}{\Gamma_D} = \frac{\Delta m_D}{\Gamma_D} =
 (0.59\pm 0.20)\% \,, \non \\
 y_D &=& \frac{ \Gamma_H - \Gamma_L}{2\Gamma_D} = \frac{\Delta \Gamma_D}{2\Gamma_D} =
 (0.80\pm 0.13)\%\,. \label{eq:data}
 \ed
Combining the errors in quadrature, the ratio of $y_D$ to $x_D$ is estimated by
 \be
 \frac{y_D}{x_D} &=& 1.36\pm 0.42\,.  \label{eq:x/y_exp}
 \ed
Intriguingly, the current data not only show $x_D \sim y_D$ but also indicate that the former is slightly smaller than the latter.

Due to the effective GIM mechanism and the absence of heavy quark enhancement,
the SD SM predictions are several orders smaller than the data \cite{Golowich:2005pt}.
It is expected that  the  GIM suppression factor might be lifted by long-distance (LD) effects \cite{HQE1,HQE2,IC,Cheng:2010rv}.
With the exclusive technique \cite{IC,Cheng:2010rv}, the results in the SM are estimated to be \cite{Cheng:2010rv}
 \be
x^{\rm SM}_D \approx (0.108 \pm 0.05)\% \,,\ \ \ y^{\rm SM}_D\approx (0.30\pm 0.34) \%\,, \label{eq:xy_SM}
 \ed
where we have averaged the possible theoretical scenarios.
Since the exclusive technique is based on the measurements of nonleptonic $D$ decays,
due to the limited accuracy of  experimental data, the SM prediction on $y_D$ is still quite uncertain.
Although the results in Eq.~(\ref{eq:xy_SM}) display the same tendency as the data,
the values of $x^{\rm SM}_D$ and $y^{\rm SM}_D$ are quite smaller than the experimental data.
Thus, the ratio in the SM is estimated as
 \be
 \frac{y^{\rm SM}_D}{x^{\rm SM}_D}= 2.78\pm 3.40 \,. \label{eq:y/x_SM}
 \ed
We see clearly that the central value by LD contributions is twice larger than that in Eq.~(\ref{eq:x/y_exp}).

If we take the central values of data in Eq.~(\ref{eq:data}) seriously, the SM results
in Eqs.~(\ref{eq:xy_SM}) and (\ref{eq:y/x_SM}) obviously cannot match with the data consistently.
For explaining the large $x_D (y_D)$ and $y_D/x_D \sim 1$, the incompatibility could be ascribed to new physics.
In most extensions of the SM, owing to the suppression of $(m_c/m_W)^2$ \cite{Chen:2007zua}
and the constraints of low-energy measurements \cite{Golowich:2006gq,Petrov:2007gp},
the SD contributions to $y_D$ with  $O(10^{-3})$ is not favorable.
Therefore, we are going to explore a peculiar new effect on the $D-\bar D$ mixing, especially on the $y_D$,
where the associated new stuff is dictated by the scale or conformal invariance and named as unparticle  \cite{Georgi1,Georgi2}.
Some interesting applications of unparticle to various systems could be referred to
Refs.~\cite{Georgi2,un1,un2,Lenz:2007nj,Cacciapaglia:2007jq,Grinstein:2008qk}.
The unique character of unparticle is {\it its peculiar phase appearing in the off-shell propagator
with positive squared transfer momentum} \cite{Georgi1}.
Due to CP invariance, the imaginary (real) part of the phase factor leads to the absorptive (dispersive)
effect of a process \cite{He:2010fz,Chen:2010rg}.
In this Letter, we investigate how $x_D$ and $y_D$ are influenced by the phase factor.
Furthermore, in order to make the production of scale invariant stuff be efficient at Large Hadron Collider (LHC),
we will concentrate on the unparticle that carries the color charges of $SU(3)_c$ symmetry \cite{Cacciapaglia:2007jq}.

Since there is no well established approach to give a full theory for unparticle interactions,
we study the topic from the phenomenological viewpoint. In order to avoid fine-tuning the parameters
for  flavor changing neutral currents (FCNCs) at tree level, we assume that the
unparticle only couples to the third generation of quarks before
electroweak symmetry breaking. Hence, the interactions obeying the
SM gauge symmetry are expressed by
 \be
 \frac{1}{\Lambda^{d_U}_U}\left[ \l_R \bar q'_R \ga_\mu  T^a q'_R \partial^\mu {\cal O}^{a}_U +
 \l_L \bar Q_L \ga_\mu  T^a Q_L \partial^\mu {\cal O}^a_U\right]\,,
\label{eq:lag_un}
 \ed
where $\l_{R, L}$ are dimensionless free parameters, $q'_R=t_R$,
$b_R$, $Q^T_L=(t, b)_L$, $\{T^a\} = \{\lambda^a/2\}$ are the
$SU(3)_c$ generators (where $\lambda^a$ are the Gell-Mann matrices)
normalized by $tr (T^a T^b) = \delta^{ab}/2$. $\Lambda_U$ is the
scale below which the unparticle is formed, and the power $d_U$ is
determined from the effective interaction of Eq.~(\ref{eq:lag_un})
in four-dimensional space-time when the dimension of the colored
unparticle ${\cal O}^{a}_U$ is taken as $d_U$. Since we only
concentrate on the phenomena of up type quarks, the associated
interactions are formulated by
 \be
 \bar U \ga_\mu \left( {\bf X}_R P_R + {\bf X}_L P_L\right) T^a U \partial^\mu {\cal O}^a_U\,,
 \label{eq:int_un}
 \ed
where  $U^T=(u, c, t)$, ${\bf X}_{R(L)}$ is a $3\times 3 $
diagonal matrix and diag(${\bf X}_{R(L)}$)=(0, 0,
$\l_{R(L)}/\Lambda^{d_U}_{U}$). After spontaneous symmetry breaking
of electroweak symmetry, we need to introduce two unitary matrices
$V^{R, L}_U$ to diagonalize the mass matrix of up type quarks. In
terms of physical eigenstates and using the equations of motion, the
interactions for $c-u-{\cal O}^a_U$ could be written as
 \be
 {\cal L}_{cu{\cal O}^a_U}&=& \frac{m_c}{\Lambda^{d_U}_{U}} \bar u \left( g^R_{uc} P_L + g^L_{uc} P_R\right)
 T^a c{\cal O}^a_U+h.c.\,, \label{eq:int_bq}
 \ed
where  the mass of light quark has been neglected. And
$g^{\chi}_{uc} = \lambda_{\chi}(V^{\chi}_U)_{13} (V^{\chi^*}_U)_{23}$
with $\chi=R, L$, in which  the index of Arabic numeral (1, 2, 3) stands for $(u, c, t)$ quark, respectively.

By following the scheme shown in Ref.~\cite{Grinstein:2008qk}, the propagator of the colored scalar unparticle is written as
 \be
\int d^4x e^{-ik\cdot x} \la 0|T {\cal O} ^{a}(x) {\cal
O}^{b}(0)|0\ra  = i \frac{C_S \delta^{ab}}{(-k^2
-i\epsilon)^{2-d_U}} \label{eq:prop_un}
 \ed
with
 \be
 C_S &=& \frac{A_{d_U}}{2\sin d_{U} \pi} \,, \non\\
 A_{d_U}&=& \frac{16\pi^{5/2}}{(2\pi)^{2d_U}} \frac{\Gamma(d_U +1/2)}{\Gamma(d_U-1) \Gamma(2d_U)}\,.
 \ed
Combining Eqs.~(\ref{eq:int_bq}) and (\ref{eq:prop_un}), the four
fermion interaction for $D$-meson oscillation is given by
 \be {\cal H}
&=& \frac{C_S }{2m^2_c }
\left(\frac{m^2_c}{\Lambda^2_U}\right)^{d_U} e^{-id_U \pi} \times
\left[ \bar u \left( g^R_{uc} P_L +g^L_{uc} P_R\right)T^a c \right]^2\,.
 \ed
For estimating the transition matrix elements,  we use
  \be
 \la \bar D | \bar u P_{R(L)} c \bar u P_{R(L)} c| D \ra &\approx&  -\frac{5}{24}  \xi_D  m_{D} f^2_{D} \,, \non \\
 \la \bar D | \bar u P_{R} c \bar u P_{L} c| D \ra &\approx&    \left(\frac{1}{24}+\frac{1}{4} \xi_D \right)m_{D} f^2_{D}\,, \non \\
 \la \bar D | \bar u_\alpha P_{R} c_\beta \bar u_\beta P_{L} c_\alpha | D \ra &\approx& \left(\frac{1} {8} + \frac{1}{12} \xi_D \right)  m_{D} f^2_{D}\,,\non\\
 \la \bar D | \bar u_\alpha P_{R(L)} c_\beta \bar u_\beta P_{R(L)} c_\alpha | D \ra &\approx& \frac{1}{24}  \xi_D m_{D} f^2_{D}\,,
 \ed
where  $ \xi_D=m^2_D/(m_c + m_u)^2$ and $f_{D}$ is the decay constant of $D$ meson.
As a consequence,  the dispersive and absorptive parts of $D-\bar D$ oscillation in the unparticle physics are found by
 \be
H^{U}_{12} &=& M^{U}_{12} - \frac{i} {2}\Gamma^{U}_{12} \,, \non
 \ed
where $M^{U}_{12}=  \cos(d_U\pi) h_U$ and $\Gamma^{U}_{12}=2\sin(d_U \pi) h_U$
with
 \be
 h_U&=&  \frac{C_S}{36 m^2_c}   \left(\frac{m^2_{c}}{\Lambda^2_U}\right)^{d_U} m_D f^2_{D} \times
 \left[  \left( g^{R^2}_{uc} +g^{L^2}_{uc}\right)\xi_D +2g^R_{uc} g^L_{uc} \right]\,.
 \ed

In order to  study the $x_D$ and $y_D$, we have to know their relations to  $M_{12}$ and $\Gamma_{12}$.
Following the notation in Particle Data Group (PDG) \cite{PDG10}, the  mass and rate differences of heavy and light
$D$ mesons could be formulated by
 \be
 \Delta m_D =Re(\Delta \omega_{HL})\,, \non \\
 \Delta \Gamma_D =-2 Im(\Delta\omega_{HL}) \label{eq:mG_D}
 \ed
with
 \be
 \Delta\omega_{HL}&=& 2\sqrt{\left(M_{12} -\frac{i}{2} \Gamma_{12}\right)\left(M^*_{12}-\frac{i}{2}\Gamma^*_{12}\right)}\,,  \label{eq:o_12}
 \ed
where $M_{12} = M^{\rm SM}_{12} + M^{U}_{12}$ and $\Gamma_{12} = \Gamma^{\rm SM}_{12} + \Gamma^{U}_{12}$.
If we define the relative phase between $M_{12}$ and $\Gamma_{12}$ to be $\phi_D= arg(M_{12}/\Gamma_{12})$,
the ratio of rate difference to mass difference is obtained by
 \be
 \frac{\Delta \Gamma_D}{\Delta m_D}&=& \frac{2r_{D} }{1- r^2_{D}/4 + R_D} \cos \phi_D
 \ed
with
 \be
 r_D&=&\frac{|\Gamma_{12}|}{|M_{12}|}\,, \non \\
 R_D&=&\sqrt{(1-r^2_D/4 )^2+r^2_D \cos\phi_D} \,.
 \ed
We note that unlike the case in $B_q$ system where the sign of $\Delta \Gamma_{B_q}$ in the SM is certain and experimental data
are consistent with SM prediction, the sign of $\Delta \Gamma_D$ in the SM is uncertain; thus we use
$\phi_D = arg(M_{12}/\Gamma_{12})$ for $D$-meson, instead of  $\phi_B = arg(-M^q_{12}/\Gamma^q_{12})$ for $B_q$-meson.
Hence, the ratio of $y_D$ to $x_D$ can be expressed by
 \be
 \frac{y_D}{x_D}=\frac{\Delta\Gamma_D}{2\Delta m_D} =\frac{r_{D}  \cos \phi_D}{1- r^2_{D}/4 + R_D} \,.
 \ed

In order to illustrate the phase effect of unparticle and simplify  the numerical estimates,
we set $\Lambda_U=1$ TeV and $g^R_{uc}= g^L_{uc} = |g_{uc}|e^{i\theta}$, $i.e.$ the couplings are vector-like.
Since the SM predictions are still quite uncertain, for numerical analysis we adopt the recent SM results to be \cite{Cheng:2010rv}
  \be
 M^{\rm SM}_{12} &=& 0.13\%\, {\rm ps^{-1}}\,, \non \\
 \Gamma^{\rm SM}_{12} &=& 0.73\%\, {\rm ps^{-1}} \,,
 \ed
where we adopt $M^{\rm SM}_{12}=  x^{\rm SM}_{D} \Gamma_D/2$ and $\Gamma^{\rm SM}_{12}=y^{\rm SM}_{D} \Gamma_D$
and we  take only the central value of $x^{\rm SM}_{D}(y^{\rm SM}_{D})$ as input. Other relevant values used
for numerical estimates are listed in Table~\ref{tab:inputs}.
\begin{table}[hptb]
\caption{ Values used for numerical estimates \cite{PDG10}. } \label{tab:inputs}
\begin{ruledtabular}
\begin{tabular}{cccc}
  $m_D$ [GeV] & $m_c$ [GeV] & $f_D$ [MeV] & $\tau_D\, [{\rm ps}]$ \\ \hline
  1.864 & 1.3 & 206.7 & 0.41 \\
\end{tabular}
\end{ruledtabular}
\end{table}

\begin{figure}[hptb]
\includegraphics*[width=5 in]{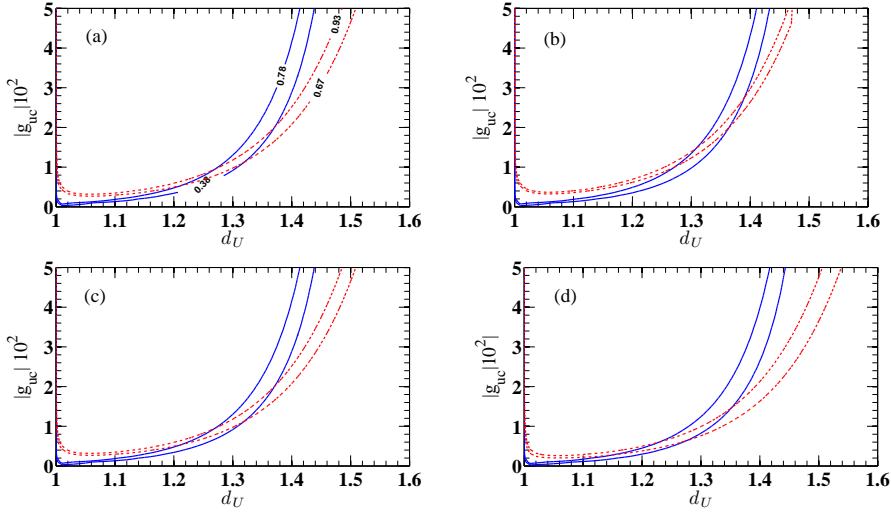}
\caption{ (a)-(d) the contours for $\Delta m_D$ (blue solid) and $\Delta\Gamma_D$ (red dotted) as a function $|g_{uc}|$
and $d_U$ with $\theta=0, \pi/4, \pi/2, 4\pi/4$, respectively. The numbers on the curves are the data with $1\sigma$ errors. }
 \label{fig:x-y}
\end{figure}

\begin{figure}[hptb]
\includegraphics*[width=4 in]{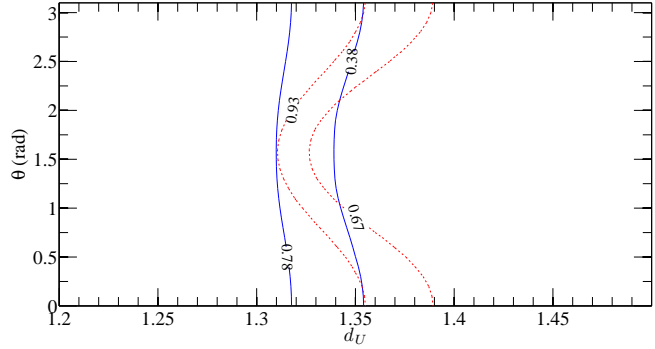}
\caption{  $x_D$ (blue solid) and $y_D$ (red dotted) as a function of $\theta$ and $d_U$ with $|g_{uc}|=1.5\times 10^{-2}$. }
 \label{fig:theta}
\end{figure}

With the chosen scenario for the free parameters and the taken numerical values,
now we have to deal with three free parameters, $i.e.$ the scale dimension $d_{U}$,
the magnitude of coupling $g_{uc}$ and its phase $\theta$.
Since the SM results are smaller than the current data, we find that the influence of $\theta$ is insignificant
when the constraints of measured $x_D$ and $y_D$ are included.
In Fig.~\ref{fig:x-y}, we present the unparticle contributions to $x_D$ and $y_D$ as a function $|g_{uc}|$
(in units of $10^{-2}$) and $d_U$, where figure (a)-(d) stands for $\theta=(0, \pi/4, \pi/2, 3\pi/4)$
and solid and dotted line denotes $x_D$ and $y_D$, respectively.
It is clear that the allowed $|g_{uc}|$ is slightly changed when $\theta$ is varied.
For further understanding the $\theta$-dependence,  we plot $x_D$ and $y_D$ as a function of $\theta$ and $d_U$
with $|g_{uc}|=1.5\times 10^{-2}$ in Fig.~\ref{fig:theta},
where the solid and dotted line corresponds to $x_D$ and $y_D$, respectively.

We have studied  the mixing parameter and lifetime difference of $D-\bar D$ oscillation in the framework of  unparticle physics,
where the new stuff is dictated by scale or conformal invariance. Unlike other models, {\it due to the peculiar phase of unparticle},
not only the mixing parameter $x_D$  but also the lifetime difference $y_D$  can be enhanced to fit the current experimental data,
especially the experimental result of  $y_D/x_D\sim 1$. We speculate that the unparticle or unparticle-like effects
could be strongly verified, when $x_D\sim y_D\sim {\rm few } \times 10^{-3}$ and $y_D/x_D \sim 1$
are satisfied simultaneously in experiments.

\begin{acknowledgments}
\noindent C.H.C would like to think Prof. Hai-Yang Cheng for useful discussions on the long-distance effects.
C.H.C was supported in part by the National Science
Council of R.O.C. under Grant No. NSC-97-2112-M-006-001-MY3.
C.S.K. was supported in part by the NRF grant funded by the Korea government (MEST)
(No. 2010-0028060) (No. 2011-0017430).

\end{acknowledgments}


\end{document}